# The Morphological Evolution of Galaxies

Roberto G. Abraham
Dept. of Astronomy & Astrophysics
University of Toronto
60 St. George Street
Toronto, Ontario M5S 3H8, Canada

Sidney van den Bergh
Dominion Astrophysical Observatory
Herzberg Institute of Astrophysics
National Research Council of Canada
Victoria, British Columbia
V9E 2E7, Canada

**ABSTRACT**

Many galaxies appear to have taken on their familiar appearance relatively recently. In the distant Universe, galaxy morphology started to deviate significantly (and systematically) from that of nearby galaxies at redshifts, z, as low as z = 0.3. This corresponds to a time ~3.5 Gyr in the past, which is only ~25% of the present age of the Universe. Beyond z = 0.5 (5 Gyr in the past) spiral arms are less well-developed and more chaotic, and barred spiral galaxies may become rarer.  By z = 1, around 30% of the galaxy population is sufficiently peculiar that classification on Hubble's traditional "tuning fork" system is meaningless. On the other hand, some characteristics of galaxies do not seem to have changed much over time. The co-moving space density of luminous disk galaxies has not changed significantly since z = 1, indicating that while the general appearance of these objects has continuously changed with cosmic epoch, their overall numbers have been conserved. Attempts to explain these results with hierarchical models for the formation of galaxies have met with mixed success.

## Introduction

Nearby galaxies are usually classified on the basis of their position on the "tuning fork" scheme originally proposed by Edwin Hubble in 1926 (*1*). This classification system is defined with reference to a set of bright nearby standard galaxies. In Hubble's scheme galaxies are divided into ellipticals and spirals. Spiral galaxies are subdivided into unbarred (S) and barred (SB) categories, which define the tines of the tuning fork. Along each tine, galaxies are further subdivided according to the tightness and fine structure of their spiral arms, which changes monotonically along the tuning fork in step with the fraction of light in the nuclear bulge of the galaxy.  These sub-categories are denoted Sa, Sb, and Sc (SBa, SBb, SBc in the case of barred spirals). A final "catch-all" category for irregular galaxies is also included.

Over 90% of *luminous* nearby galaxies fit within this system. Intrinsically faint galaxies, such as many of the dwarf galaxies in the Local Group (the agglomeration of galaxies within 1.3 Mpc of the Milky Way Galaxy), cannot be slotted into the standard classification system. Such faint dwarf galaxies vastly outnumber the bright galaxies described by the Hubble sequence. However, these dwarfs contribute little to the total mass budget of the galaxy population, and they are difficult to detect at large distances.

Modern attempts to understand the physical significance of Hubble's classification system are based on the idea that most matter in the Universe is not in stellar or gaseous form, but is instead comprised of dark matter. Dark matter does not emit or absorb radiation, and can only be detected through its gravitational effects on spectroscopically observed galaxy rotation curves, or by gravitational bending of galaxy images ("gravitational lenses"). Some fraction of the dark matter in the Universe is made up of baryons (protons and neutrons and related particles), but from the relative abundances of elements formed in the Big Bang, well over 90% must be in a non-baryonic form[1]. Theoretical work relies on the hypothesis that dark matter and galaxies are linked, because gravitating concentrations of dark matter originating soon after the Big Bang are responsible for the formation of galaxies. In this picture (Figure 1), small concentrations of dark matter grow slowly at first, being gradually compressed by self-gravity. However, when the concentrations reach a critical density (about 200 times the mean background density of the Universe), they undergo a catastrophic non-linear collapse, which results in the formation of an extended "halo" of dark matter. Over time these halos clump together under their mutual gravitational attraction, merging to form a hierarchy of larger halos. The rate of cooling of hydrogen gas drawn into these large halos governs the assembly of normal galaxies, and ultimately their morphology.

This new paradigm has transformed our view of galaxy evolution. It is now believed that most visible galaxies are embedded in much larger and more massive dark matter halos that detached from the expanding cosmic plasma (created in the Big Bang) at early times. Furthermore, galaxies are no longer viewed as growing in isolation, but rather as being linked into a web of large scale structure, which originated in the density fluctuations traced by the surface brightness variations now seen in the cosmic microwave background (*2-5*). Observations of galaxy morphology now span about 70% of the total age of the Universe and allow this paradigm to be tested.

## Caveats

Morphological classification of galaxies at redshifts near z = 1 is challenging because the number of pixels per image may be as much as ~100 times smaller than in the images of nearby galaxies. Classification at z ~ 1 therefore represents a considerable extrapolation from similar work at z ~ 0. Caution has to be exercised to avoid resolution-dependent effects that might affect images of distant galaxies more than they do nearer galaxies. The slight "under sampling" of images on HST's Wide-Field/Planetary Camera 2 makes the classification of very compact galaxies (such as distant ellipticals) particularly difficult. Another problem is that it is often tempting to "shoehorn" slightly peculiar distant galaxies into the familiar Hubble classification system. As a result, one classifier may regard a certain object as, say, a somewhat peculiar variant of a regular spiral galaxy (designated "Spec" in many catalogs of local galaxies, and considered to lie within the Hubble sequence), while another classifier may classify the same object as peculiar (i.e. outside the Hubble sequence altogether). Such taxonomical inexactitude is probably inherent in the subjective nature of visual classification, and is the major reason for the rising popularity of computer-based objective classification schemes based on the measurement of quantitative parameters. However, such objective schemes do not yet encompass the full richness of galaxy forms. Computer-based classification systems presently only allow one to group galaxies into broad categories that most visual morphologists would consider too coarse (*6, 7*). For example, no quantitative scheme can currently distinguish between physically distinct sub-categories of spiral structure[2].

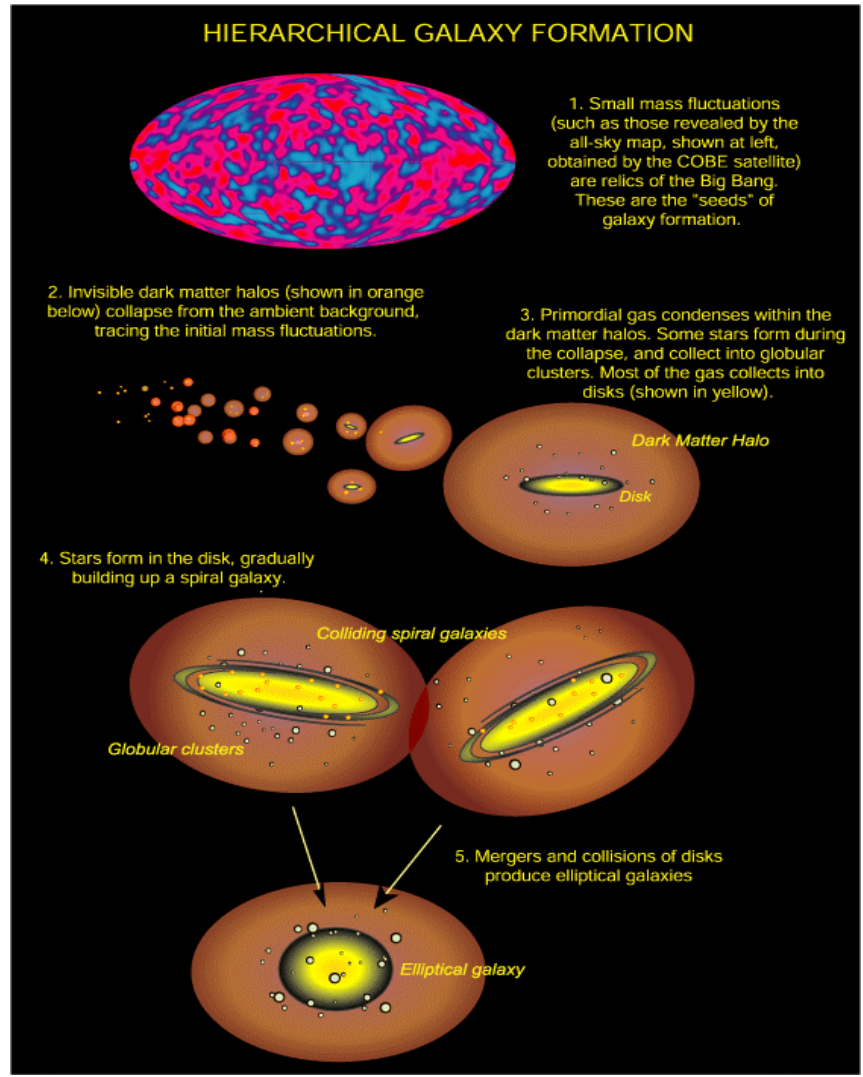

Figure 1: Schematic illustration of the formation of galaxies in the hierarchical formation paradigm. The timescale for the various phases shown in the illustration depends on cosmological parameters (i.e. the value of the Hubble constant, and the total mass-energy of the Universe contained in both dark matter and dark energy). In an accelerating Universe most elliptical galaxies form at redshifts z>1, while in a decelerating Universe most ellipticals form at z<1.

Another important caveat that applies to all current studies of distant galaxy morphology is that the number of galaxy images available for study is small. While each Hubble Deep Field contains around 3000 galaxies or galaxy fragments, only a few hundred of these are both large enough, bright enough, and near enough to allow reliable morphological classification, and of these only ~70 are sufficiently well-resolved (and are viewed at sufficiently low-inclinations to the line-of-sight) to allow substructure such as bars and weak spiral arms (if present) to be detected. Also, as a result of the morphology-density relation (9), narrow-field observations such as those undertaken with HST may be significantly biased by galaxy environment. For example, elliptical galaxies are expected to be overabundant in those parts of the line-of-sight that pass through, or close to, clusters of galaxies.

A final complication is introduced by the fact that distant galaxies are seen at a range of redshifts. Therefore observations made in a single filter probe each galaxy at a wavelength different from that of the same filter in the rest frame. This poses a problem for morphologists, because the appearance of a galaxy can vary strongly with wavelength. In the ultraviolet a galaxy's light comes mainly from hot young stars, which are often distributed in clumpy irregular knots of star formation. At optical wavelengths stars on the stellar main sequence dominate, and galaxies take on their most familiar appearance. In the infrared most of the flux is usually from evolved old stars, which are quite uniformly distributed, so galaxies appear smoother than at optical wavelengths. As a result, identical galaxies appear different at different redshifts. The importance of this effect (known as the "morphological K-correction") depends rather strongly upon the distance at which the galaxy is seen.

For galaxies with redshifts z<0.8 (i.e. for galaxies seen at look-back times less than about halfway back to the Big Bang[3]), observations made with HST using the F814W filter (commonly used for morphological work) correspond to observations made at familiar optical wavelengths in the galaxy's frame of reference. Thus the effects of morphological K-corrections are quite benign at z<0.8, and much work on galaxy morphology has focused on this "safe" redshift range. At redshifts z>0.8, galaxies viewed with the F814W filter are being seen at ultraviolet wavelengths in the rest frame. Unfortunately, rather little is known about the appearance of nearby galaxies in the ultraviolet, because our atmosphere is opaque to UV radiation, and only rather small UV imaging surveys have been undertaken from space. It is clear that some nearby galaxies that appear optically normal appear strange in the UV (*10, 11*). Consequently galaxies at z>0.8 are best studied in the infrared at wavelengths greater than 1μm in order to undertake fair comparisons with local studies made at optical wavelengths (*12*). HST had only a rather limited capability to undertake such observations, using the short-lived NICMOS camera.

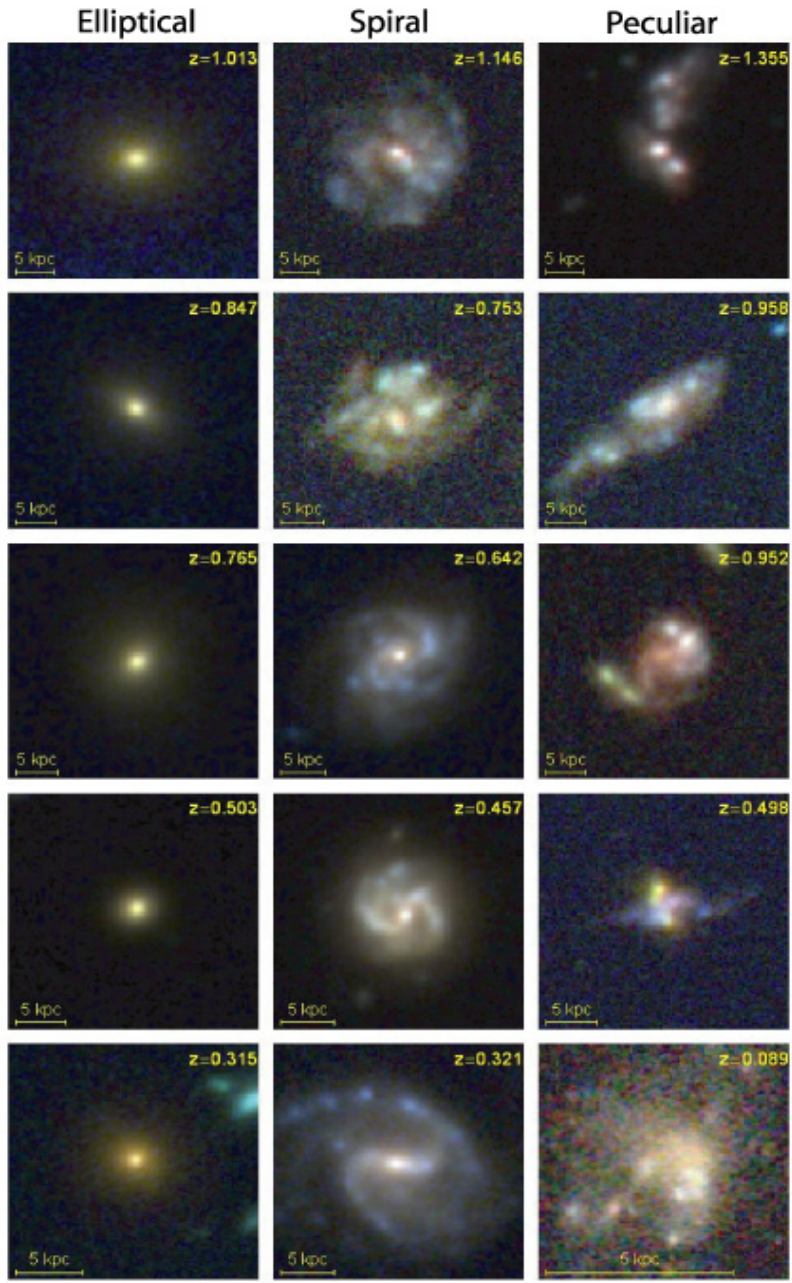

Figure 2: Montage of typical elliptical (left column), spiral (middle column), and peculiar (right column) galaxies in the Northern and Southern Hubble Deep Fields, imaged by the Hubble Space Telescope. The images are sorted by redshift (local galaxies at bottom, distant galaxies at top). Color images were constructed by stacking images obtained through blue, yellow, and near-infrared filters. Note the gradual loss in the organization (and the increase in the fragmentation) of spiral arms in the spirals. Barred spirals (such as the example shown in bottom row) become rare beyond z=0.5. The physical nature of most of the peculiar galaxies in the far right column is poorly understood. The exception is the object at the bottom-right corner, which is a dwarf irregular, a class of objects common in the local Universe. In Figure 4 we speculate on the possible nature of two other galaxies in the far-right column.

## Morphological Classifications of Distant Galaxies

The existence of a population of morphologically peculiar galaxies at high redshifts was first suspected on the basis of HST observations as early as 1994 (*13-17*). Much of this early work proceeded hand-in-hand with improved attempts to quantify galaxy morphology using automated morphological classification schemes, as fast and robust replacements for traditional visual classifications. Present state-of-the-art automatic classification systems can group galaxies in broad categories (such as spirals, ellipticals, and peculiars) as reliably as can a visual classifier (*6, 18, 19*), and these systems are in principle extensible to encompass wholly new classes of galaxies beyond those included within the tuning fork paradigm.

While these early HST studies suggested that extending the low-redshift tuning fork paradigm to high redshifts would prove difficult, the ultimate futility of this task only became apparent when the Hubble Deep Field images were released (Figure 2). At the limit to which morphological classification is reliable in the deep fields (variously quoted as between I=25 mag and I=26 mag) around 30% of the galaxies are classified as "strongly peculiar or merging", i.e. about 30% of the galaxies fall outside the conventional framework (*18, 20, 21*). Most studies find little evidence for evolution in the space density of elliptical galaxies (with the exception of (*22*)). But, as is the case with spiral galaxies, the internal characteristics of elliptical galaxies do show trends with redshift. The most prominent of these is an increasing proportion of elliptical galaxies with blue cores in the distant Universe (*23*), which indicate the presence of some hot young blue stars in these old galaxies.

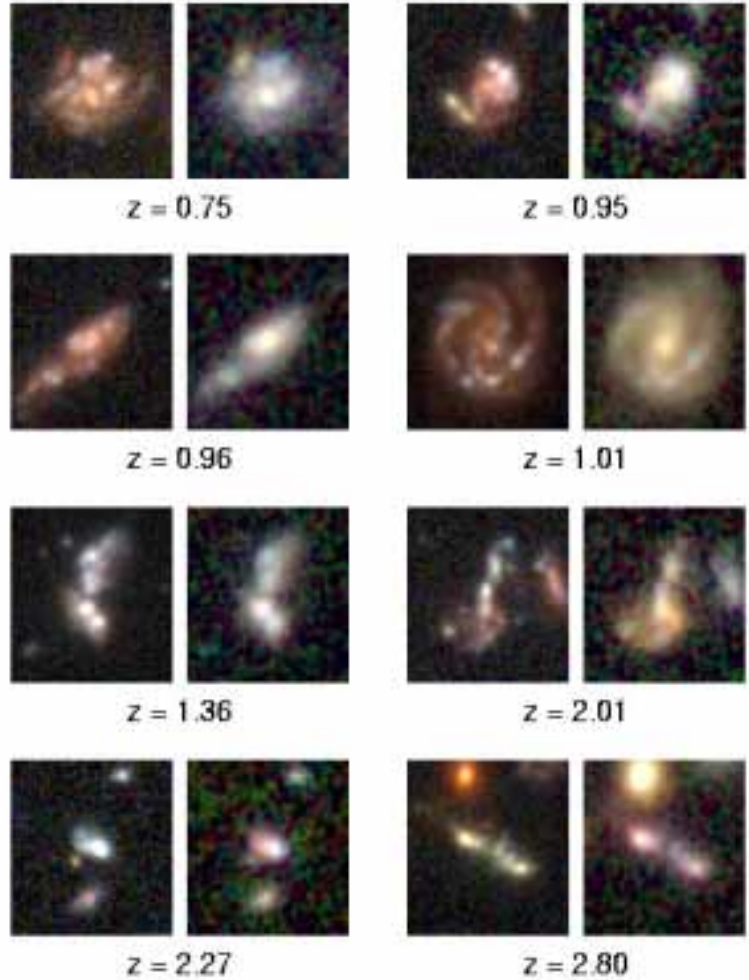

*Figure 3:* A montage, taken from Ferguson, Dickinson and Williams (2000), comparing the optical and infrared morphologies of selected galaxies at various redshifts from the northern Hubble Deep Field. "True color" composites in the left-hand panels are made using a blue channel (central wavelength=450nm), a "visual" channel (central wavelength=606nm), and a channel straddling the optical/near-IR border (central wavelength=814nm). The right hand panels are constructed using IR channels (central wavelengths of 814nm, 1100nm, and 1600nm), and have slightly poorer resolution than the corresponding images at left. For most objects, including the most distant galaxies, the optical and near-IR morphologies are similar, indicating that the morphological peculiarities are intrinsic to the galaxies, and are not the result of the changing rest-wavelength of the observations. [Courtesy H. Ferguson, M. Dickinson, R. Williams, STScI and NASA].

Imaging studies allow trends to be established as a function of galaxy brightness, but more physically based investigations of galaxy morphology require that the redshifts of the galaxies be known. This is a formidable challenge; galaxies can be detected on CCD images down to much fainter levels than can be probed spectroscopically, because in the latter case light is being dispersed, so Poisson noise is larger relative to the signal of interest. One rather successful technique for dealing with this difficulty has been to estimate redshifts from galaxy colors (*24*). Morphological investigations using these "photometric redshifts" (*22, 25*) have confirmed and solidified the statistical conclusions from earlier imaging studies of galaxies that had no photometric redshift information, and allowed samples of objects at be constructed over redshift ranges immune to the effects of morphological K-corrections. But because photometric redshift techniques rely, at a basic level, on assumptions regarding the possible spectra of galaxies and on the nature of systematics that may alter such spectra (such as the form of the dust extinction curve), they are only fairly crude estimators ($\Delta z \sim 0.1$ for blue galaxies, and $\Delta z \sim 0.04$ for red galaxies) of the true redshifts of distant galaxies (*24*). Morphological investigations of large samples of objects of known redshift became feasible when Brinchmann et al. (*26*) obtained Hubble Space Telescope images for 341 galaxies of known redshift from the Canada-France Redshift Survey. Results from this investigation agree with earlier results (*15, 18, 27*). Around 30% of galaxies at z=1 are morphologically peculiar, and these authors found that the fraction of galaxies with irregular morphology increases with redshift beyond the range that is expected on the basis of the systematic misclassifications of spiral galaxies introduced by band-shifting effects. Furthermore, their data for the frequency with which elliptical galaxies are distributed over the redshift range $0.0 < z < 1.2$ were found to be consistent with expectations for scenarios with little or no morphological evolution for these objects.

The trends described so far have been based on rather coarse automated morphological classifications. More fine-grained "precision morphology" (which can subdivide spiral galaxies into various classes, for example) still requires a trained human eye, and also requires complete redshift information and multi-filter data in order to synchronize data to a uniform rest wavelength. The largest such investigation (*28*) has been based on the Caltech Faint Galaxy Redshift Survey (CFGRS), a deep follow-up redshift survey centered on the Northern Hubble Deep Field, but also containing redshifts for galaxies in the nearby HDF "flanking fields". This study suggests that: (i) The fraction of early-type (E - Sab) galaxies remains approximately constant over the range $0 < z < 1$, in agreement with earlier studies (*26*)[4]. (ii) The fraction of intermediate/late- type (Sb - Ir) galaxies drops by a factor of ~2 over this redshift range, presumably because many late-type galaxies viewed at a look-back times of ~8 Gyr have been classified as "peculiar" or "merger". (iii) The fraction of such peculiar or merging galaxies exhibits a monotonic increase from 5% at z ~0, to 10% at z ~0.4, 19% at z ~0.7 and 30% at z ~0.95. These numbers are in quite good agreement with the corresponding numbers inferred from measurements based on automated classifications, though it should be once again emphasized that the numbers of galaxies in all such studies are small, and the conclusions which can drawn from them are correspondingly uncertain. It is worth noting that the merger rate estimated from the HDF on the basis of morphological considerations is in reasonable agreement with the merger rate inferred from HST imaging of the CFRS fields (*30*), and from pair counts at lower redshifts (*31*).

Beyond z~1, one must observe galaxies in the infrared in order to study them at familiar optical wavelengths in their rest frames. The NICMOS camera on HST had this capability, although its small field of view and limited lifetime[5] meant that relatively few z>1 field galaxies were observed in the near infrared before the instrument ceased functioning. However, over the short lifetime of this camera it was possible to demonstrate (*12, 32, 33*) that *most galaxies with peculiar optical morphologies remain peculiar when viewed in the near infrared*, at least out to z=3 (Figure 3). It follows that the strange appearance of these objects is not merely the consequence of "morphological K-corrections" (i.e. of observing only irregularly distributed sites of star formation at rest ultraviolet wavelengths, while missing the bulk of the galaxy). The peculiar appearance of these objects reflects a genuinely irregular structural state in these galaxies. However, in the most distant objects (i.e. at z>3), it may be the case that we really are only seeing the very tip of the baryonic mass iceberg (i.e. only the youngest stars in the galaxies, grouped together in irregular star formation complexes).

Returning our consideration to the luminous peculiar galaxies seen at lower redshifts (1<z<3), it is fair to say that little is known about the nature of these objects. A large fraction may be galaxy mergers or collisions, because the abundance of colliding galaxies is expected to increase in the distant Universe[6], although large-sample statistical analyses of the merger rate as a function of cosmic epoch at z>1 have not been undertaken. Most objects classified as "mergers" in deep HST images are only merger candidates. Radial velocity differences of less than a few hundred km s$^{-1}$, and/or evidence for tidal distortions (such as tails), are required to be sure that merger candidates are indeed physically interacting/merging objects. However, the evidence for mergers is at least strongly suggestive, since some of the strangest-looking systems can be matched to counterparts amongst nearby merging systems (Figure 4). But even if most of the high redshift peculiar galaxies turn out to be distorted due to mergers or collisions, this is probably not the whole story, since some peculiar systems exhibit properties (such as synchronized internal colors) that are not seen in nearby mergers (*31*).

Two types of observations are needed understand the nature of the morphologically peculiar galaxy population. Firstly, rest-frame infrared imaging is needed in order to look for the presence of red light from evolved stars in these galaxies. Such data will constrain the ages of the underlying stars in the galaxies independently from their morphologies. Such data will also determine whether the objects are genuinely proto-galactic (i.e. formed exclusively from young stars), or whether they correspond to a stage in the growth of evolved galaxies. Secondly, dynamical studies of high redshift objects are needed in order to distinguish merger remnants from *bona fide* singular new classes of objects. Such studies are in their infancy. On the basis of early work on the so-called "Lyman break" population at z>2.3, much of which is morphologically peculiar, it seems that at least some of the more distant morphologically peculiar systems do not exhibit even the most basic ordered motions expected from rotating disks (*34*). Some of these objects could be distant analogs to the most extreme local star-bursting galaxies, whose morphological characteristics originate in the collapse along filaments of gas driven by energetic "super winds" (*35-37*).

## Comparison with Numerical Simulations

The computational state-of-the-art has evolved to the point where numerical simulations of galaxy evolution can be linked with cosmological simulations for the growth of large-scale structure (such as galaxy clusters). Consequently the morphological evolution of galaxies can now be explored in a proper cosmological context. However, the resolution of such simulations remains rather poor (< $10^5$ particles per galaxy). Simulations with higher resolution have been undertaken with pure n-body gravitational codes, at the expense of neglecting important gas-dynamical effects. Alternatively, simulations can be constructed to study individual galaxies, or small systems of galaxies, essentially in isolation. Such simulations may be constructed with up to $10^8$ particles per galaxy, i.e. within a factor of 100 of the number of stars in a typical luminous galaxy. This class of simulations suggests that elliptical galaxies could form from the collision of spirals (*38, 39*). An oft-cited prediction of hierarchical models is that elliptical galaxies should become increasingly rare with increasing redshift, dropping to 30% their local co-moving space density by z~1. This evolution can probably be excluded on the basis of existing data (though see (*22*) for a contrary opinion). However, models which include a cosmological constant now push the epoch for most elliptical formation to z>1, which is beyond the redshift at which an unbiased inventory of ellipticals exists. Therefore merger-induced formation of field ellipticals remains the favored theoretical model, although this scenario still needs observational verification.

The present generation of n-body/gas-dynamical models fails to account for a number of important observations. Dark halos appear to have constant density cores, rather than the predicted power law-like cusps, and the models predict an order of magnitude too many companions for giant galaxies. However, the most significant problem is their failure to correctly account for galaxy sizes. To date *all* cosmological codes (i.e. those that begin with a fluctuation spectrum for dark matter halos at high redshift, and evolve this forward while collapsing gas into these halos to form galaxies) predict that luminous galaxy disks around z~1 should be far smaller than are typically seen. The observational constraint is actually rather severe: the space density of large disks at z~1 is similar to that seen in the nearby Universe (*40*). "Semi-analytical" models (which blend the predictions of n-body codes for the growth of dark matter halos with phenomenological models for the growth of galaxies within these halos) are more successful in predicting galaxy sizes as a function of redshift, as well as in modeling the changing morphological mix with redshift. A criticism of these models is that they adopt rather crude scaling relations to describe important aspects of physics that cannot be treated from first principles (e.g. stellar winds expelled from the disks of galaxies which lead to star formation by a complex feedback process).

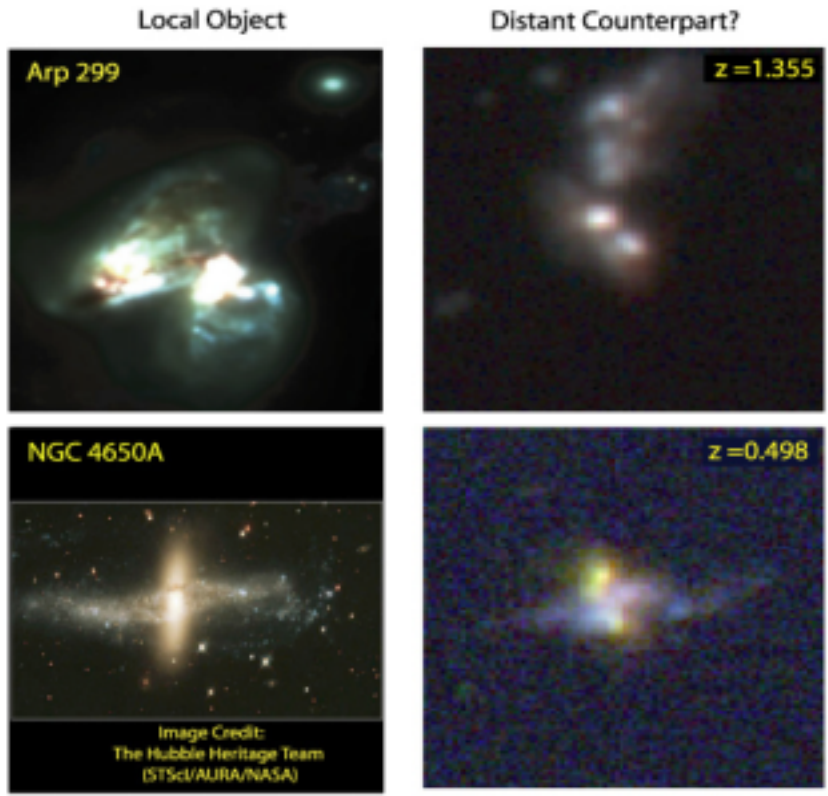

Figure 4: Some speculations on the nature of high-redshift peculiar galaxies. [Top row] As pointed out by Hibbard & Vacca (1997), the nearby local merging starburst galaxy Arp 299 shares many of the morphological and star-formation of characteristics of "chain-like" peculiar galaxies, such as the example (HDFN_J123652.78+621354.3) shown at right. While the correspondence is not exact (chain-like peculiars exhibit more components than is typical seen in local starbursts, and sometimes show a curious synchronization of internal colors), local starbursts remain the closest match to many distant systems. [Bottom row] Classes of merging galaxies that are exceedingly rare in the local Universe may be responsible for some of the oddest systems seen in the Hubble Deep Fields. For example, the "polar ring" galaxy NGC 4650A bears a close resemblance (both in terms of morphology and in terms of relative color) to the distant galaxy (HDFN_J123652.88 + 621404.8) shown at bottom right. Galaxy collisions can produce a rich variety of morphologically peculiar galaxies, many of which do not resemble "textbook" mergers, i.e. a pair of objects with obvious tidal tails.

In summary, at present the numerical models that attempt to model the physics of galaxy formation from first principles do not have sufficient resolution, while those that attempt to sidestep problems with resolution oversimplify the systems. At present no class of models has accounted for detailed evolution of the internal structures of galaxies. An interesting link between dynamical models and morphology is the predicted evolution of bars, since these features are sensitive to the dynamical state of galaxy disks and are bright enough to be detectable in distant galaxies. In an inspection of the Northern Hubble Deep Field images, van den Bergh suggested that barred spirals were unexpectedly rare at high redshifts (*17*). Support has been lent to this observation by (*41*), which shows that the frequency of SB galaxies at 0.6<z<0.8 is only ~5% in the Hubble Deep Fields, as compared to 21-34% among nearby Shapley-Ames galaxies. Both of these results are based on small numbers of galaxies, and need to be tested against larger samples.[7]

## Discussion

Table 1 provides a summary of our interpretation of the "key ages" in the morphological development of galaxies to z=1. Beyond this redshift, the galaxy population appears so peculiar that we feel much more infrared imaging data (probing galaxies at optical wavelengths in the rest frame) is needed in order to draw any firm conclusions beyond the basic observation that eight billion years ago luminous galaxies looked very different from how they look today. Perhaps the most surprising result from Table 1 is that the systematics of galaxy morphology still appear to be evolving rapidly at the present time. This is most clearly illustrated by the fact that grand-design spirals are found to be rare before z ~ 0.3, corresponding to a look-back time of only 3.5 Gyr. By the time that one looks back beyond z = 0.5 (5 Gyr) almost all spiral structure seems to be less well-developed (or more chaotic) than it is among nearby galaxies.

Explaining why the Hubble classification scheme appears to "evaporate" at redshifts z > 0.3 is a challenge that can only be met by a greater understanding of galaxy dynamics at high redshifts. The gas fraction in typical spiral galaxies may have decreased by a large fraction between z = 1 and z = 0, and encounters with giant molecular clouds, spiral arms and perhaps clumps in the dark matter halo may have increased the velocity dispersion within the disk at high redshifts. Fuchs & von Linden (1998) and Fuchs et al. (2000) suggest that either of these effects would impact the stability of spiral disks in the distant past, but the data simply do not yet exist to allow astronomers to determine whether either of these effects plays a dominant role. At this stage all that is clear is that the morphology of spiral galaxies is evolving rapidly, and systematically, even at quite low redshifts. Familiar types of galaxies, such as barred and grand-design spirals, appear to be relatively recent additions to the extragalactic zoo. The nature of the many morphologically peculiar galaxies at high redshift remains a complete mystery. These objects might be mergers, proto-galaxies, new classes of evolved systems, or a combination of all three.

We are only beginning to glimpse the morphology of galaxies in the distant Universe, and are groping to fit them into a physically meaningful organizational structure. Progress will require astronomers to devise better (and more fully quantitative) means of characterizing the (increasing) peculiarity of galaxies in the distant Universe. As a first step one might follow in the footsteps of Morgan (*43, 44*), Doi et al. (*45*), and Abraham (*17, 46*), who used central concentration of light as a classification parameter. Other statistical measures that could turn out to be useful for the classification of galaxies at large look-back times are rotational asymmetry (*18, 47*), "blobbiness" and filling factor of isophotes (*48*), and central offset of the galaxy nucleus. Many of these are likely to be correlated with both the ages and star-formation properties of galaxies. The physical significance of these parameters can be better understood by explicit reference to n-body/gas-dynamical simulations. With the infrared-optimized *Next Generation Space Telescope* on the horizon, it will also be of great interest to extend rest-wavelength optical morphological work to distant objects at ever greater lookback times that can best be observed in the infrared (*49*). Extending morphological studies to the rest-frame infrared has a number of important benefits: dust obscuration is reduced, and the link between morphology and underlying stellar mass (best traced by old stars) becomes more direct.

It seems conceivable that in some of the distant morphologically peculiar objects in the Hubble Deep Fields we are witnessing an early stage in the life cycle of present-day luminous galaxies, in which most of their baryons become locked up in stars, rather than in primordial gas. This is the classic description of a protogalaxy. Ground-based spectroscopic surveys show that over the redshift range 0<z<4 most star-formation is occurring in massive galaxies (*50, 51*). However, by z=10 the age of the Universe is rapidly approaching the dynamical timescale of a large galaxy (~ few $10^8$ years), while stellar evolutionary timescales for massive stars remain much shorter ($10^4$-$10^7$ years). Most of the stars that exist at these high redshifts may not have had time to come together to form large galaxies. Indeed, deep HST images are littered with small faint objects that might be more properly considered galaxy fragments (*52, 53*). Their distances are unknown; only the brightest 5% of objects on the Hubble Deep Fields have known redshifts. It is conceivable that these systems – and not the morphologically bizarre, luminous galaxy population that has been the focus of much of this review – represent the true first appearance of galaxies in the Universe. In any case, these faint galaxies are barely resolved, and we have no idea what they look like; they are both too faint and too small to classify with HST. They remain outside the realm of the morphologist. At present they define, in the words of Hubble, the "…dim boundary — the utmost limits of our telescopes. There, we measure shadows, and we search among ghostly errors of measurement for landmarks that are scarcely more substantial" (*54*).

## Acknowledgements
We thank Richard Ellis, Jarle Brinchmann, Michael Merrifield, Roger Blandford, Judith Cohen, and David Hogg for most enjoyable collaborative investigations into high-redshift galaxy morphology. Thanks are also due to Jerry Sellwood for sharing his thoughts on the formation and destruction of bars

**Table 1:** Summary of key ages in galaxy morphology to z=1

| Redshift | Look-back Time | Key Developments in Galaxy Morphology |
|---|---|---|
| z<0.3 | <~3.5 Gyr | Grand-design spirals exist. Hubble scheme applies in full detail. |
| z~0.5 | ~5 Gyr | Barred spirals become rare. Spiral arms are underdeveloped. The bifurcated "tines" of the Hubble tuning fork begin to evaporate. |
| z>0.6 | >6 Gyr | Fraction of mergers and peculiar galaxies increases rapidly. By z=1 around 30% of luminous galaxies are off the Hubble sequence. |

---

[1] In the last three years astronomers have also found evidence for "dark energy", a counterpart to dark matter that may be responsible for an accelerating expansion of the Universe

[2] Quantitative investigations of spiral structure at high redshifts may become possible in the near future, however, using Fourier decomposition techniques (*8*).

[3] All ages used in the present paper are based on a cosmology in which the Hubble parameter is $H_0 = 70$ km s$^{-1}$ Mpc$^{-1}$, the contribution of baryonic and dark matter to the critical closure mass-energy density of the Universe is $\Omega_M=0.3$, and the contribution of dark energy is manifested as a cosmological constant with density is $\Omega_\Lambda= 0.7$. In this cosmology the present age of the Universe from the time of the Big Bang is 13.5 Gyr.

[4] It should be emphasized that around 30% of the objects classed as ellipticals at high redshifts do not have the uniform red colors of local elliptical galaxies. Perhaps some of the high-redshift "blue ellipticals" are related to a class of objects known as "blue compact galaxies" in the nearby Universe (*29*).

[5] NICMOS was available for a two-year window of observations. After a three-year hiatus NICMOS will once again become available after the next HST servicing mission.

[6] Since the proper volume between galaxies diminishes with redshift as $(1+z)^{-3}$, distant galaxies are packed more tightly into a given volume of space.

[7] It is worth noting that the absence of bars cannot be due to morphological K-corrections (as suggested in (*42*)), since it is seen to occur at redshifts in which the HST observations are well synchronized to blue optical wavelengths, i.e. where at rest wavelengths where we are most familiar with the morphologies of nearby galaxies. Similarly, the effect cannot be due to low signal-to-noise in distant galaxy images, since simulations show that over 90% of the strongly barred nearby galaxies in the *Ohio State University Bright Spiral Galaxy Survey* would remain obviously barred if seen at z=0.7 in the Hubble Deep Field (van den Bergh et al. 2001, in preparation).